\documentclass[10pt,a4paper,onecolumn]{article}
\usepackage{marginnote}
\usepackage{graphicx}
\usepackage{xcolor}
\usepackage[utf8]{inputenc}
\usepackage{authblk,etoolbox}
\usepackage{titlesec}
\usepackage{calc}
\usepackage{tikz}
\usepackage{hyperref}
\hypersetup{colorlinks,breaklinks=true,
            urlcolor=[rgb]{0.0, 0.5, 1.0},
            linkcolor=[rgb]{0.0, 0.5, 1.0}}
\usepackage{caption}
\usepackage{tcolorbox}
\usepackage{amssymb,amsmath}
\usepackage{ifxetex,ifluatex}
\usepackage{seqsplit}
\usepackage{xstring}

\usepackage{float}
\let\origfigure\figure
\let\endorigfigure\endfigure
\renewenvironment{figure}[1][2] {
    \expandafter\origfigure\expandafter[H]
} {
    \endorigfigure
}

\usepackage{fixltx2e} 
\usepackage[
  backend=biber,
]{biblatex}
\bibliography{paper.bib}

\let\textttOrig=\texttt
\def\texttt#1{\expandafter\textttOrig{\seqsplit{#1}}}
\renewcommand{\seqinsert}{\ifmmode
  \allowbreak
  \else\penalty6000\hspace{0pt plus 0.02em}\fi}

\makeatletter
\let\href@Orig=\href
\def\href@Urllike#1#2{\href@Orig{#1}{\begingroup
    \def\Url@String{#2}\Url@FormatString
    \endgroup}}
\def\href@Notdoi#1#2{\def\tempa{#1}\def\tempb{#2}%
  \ifx\tempa\tempb\relax\href@Urllike{#1}{#2}\else
  \href@Orig{#1}{#2}\fi}
\def\href#1#2{%
  \IfBeginWith{#1}{https://doi.org}%
  {\href@Urllike{#1}{#2}}{\href@Notdoi{#1}{#2}}}
\makeatother

\newlength{\cslhangindent}
\newlength{\csllabelwidth}
\newenvironment{CSLReferences}[3] 
 {
  \setlength{\parindent}{0pt}
  \ifodd #1 \everypar{\setlength{\hangindent}{\cslhangindent}}\ignorespaces\fi
  \ifnum #2 > 0
  \setlength{\parskip}{#2\baselineskip}
  \fi
 }%
 {}
\usepackage{calc}

\usepackage[top=3.5cm, bottom=3cm, right=1.5cm, left=1.0cm,
            headheight=2.2cm, reversemp, includemp, marginparwidth=4.5cm]{geometry}

\titleformat{\section}
  {\normalfont\sffamily\Large\bfseries}
  {}{0pt}{}
\titleformat{\subsection}
  {\normalfont\sffamily\large\bfseries}
  {}{0pt}{}
\titleformat{\subsubsection}
  {\normalfont\sffamily\bfseries}
  {}{0pt}{}
\titleformat*{\paragraph}
  {\sffamily\normalsize}

\usepackage{fancyhdr}
\makeatletter
\let\ps@plain\ps@fancy
\fancyheadoffset[L]{4.5cm}
\fancyfootoffset[L]{4.5cm}

\definecolor{linky}{rgb}{0.0, 0.5, 1.0}

\newtcolorbox{repobox}
   {colback=red, colframe=red!75!black,
     boxrule=0.5pt, arc=2pt, left=6pt, right=6pt, top=3pt, bottom=3pt}

\patchcmd{\@maketitle}{center}{flushleft}{}{}
\patchcmd{\@maketitle}{center}{flushleft}{}{}
\patchcmd{\@maketitle}{\LARGE}{\LARGE\sffamily}{}{}
\def\maketitle{{%
  
  \AB@maketitle}}
\makeatletter
\renewcommand\AB@affilsepx{ \protect\Affilfont}
\renewcommand\AB@affilnote[1]{{\bfseries #1}\hspace{3pt}}
\renewcommand{\affil}[2][]%
   {\newaffiltrue\let\AB@blk@and\AB@pand
      \if\relax#1\relax\def\AB@note{\AB@thenote}\else\def\AB@note{#1}%
        \setcounter{Maxaffil}{0}\fi
        \begingroup
        \let\href=\href@Orig
        \let\texttt=\textttOrig
        \let\protect\@unexpandable@protect
        \def\thanks{\protect\thanks}\def\footnote{\protect\footnote}%
        \@temptokena=\expandafter{\AB@authors}%
        {\def\\{\protect\\\protect\Affilfont}\xdef\AB@temp{#2}}%
         \xdef\AB@authors{\the\@temptokena\AB@las\AB@au@str
         \protect\\[\affilsep]\protect\Affilfont\AB@temp}%
         \gdef\AB@las{}\gdef\AB@au@str{}%
        {\def\\{, \ignorespaces}\xdef\AB@temp{#2}}%
        \@temptokena=\expandafter{\AB@affillist}%
        \xdef\AB@affillist{\the\@temptokena \AB@affilsep
          \AB@affilnote{\AB@note}\protect\Affilfont\AB@temp}%
      \endgroup
       \let\AB@affilsep\AB@affilsepx
}
\makeatother

\renewcommand\Affilfont{\sffamily\small\mdseries}
\ifnum 0\ifxetex 1\fi\ifluatex 1\fi=0 
  \usepackage[T1]{fontenc}
  \usepackage[utf8]{inputenc}

\else 
  \ifxetex
    \usepackage{mathspec}
    \usepackage{fontspec}

  \else
    \usepackage{fontspec}
  \fi
  \defaultfontfeatures{Ligatures=TeX,Scale=MatchLowercase}

\fi
\IfFileExists{upquote.sty}{\usepackage{upquote}}{}
\IfFileExists{microtype.sty}{%
\usepackage{microtype}
\UseMicrotypeSet[protrusion]{basicmath} 
}{}

\usepackage{hyperref}
\hypersetup{unicode=true,
            pdftitle={LightSAFT: Lightweight Latent Source Aware Frequency Transform for Source Separation},
            pdfborder={0 0 0},
            breaklinks=true}
\usepackage{longtable,booktabs}

\let\addcontentslineOrig=\addcontentsline
\def\addcontentsline#1#2#3{\bgroup
  \let\texttt=\textttOrig\addcontentslineOrig{#1}{#2}{#3}\egroup}
\let\markbothOrig\markboth
\def\markboth#1#2{\bgroup
  \let\texttt=\textttOrig\markbothOrig{#1}{#2}\egroup}
\let\markrightOrig\markright
\def\markright#1{\bgroup
  \let\texttt=\textttOrig\markrightOrig{#1}\egroup}

\usepackage{graphicx,grffile}
\makeatletter
\def\maxwidth{\ifdim\Gin@nat@width>\linewidth\linewidth\else\Gin@nat@width\fi}
\def\maxheight{\ifdim\Gin@nat@height>\textheight\textheight\else\Gin@nat@height\fi}
\makeatother
\setkeys{Gin}{width=\maxwidth,height=\maxheight,keepaspectratio}
\IfFileExists{parskip.sty}{%
\usepackage{parskip}
}{
\setlength{\parindent}{0pt}
\setlength{\parskip}{6pt plus 2pt minus 1pt}
}
\ifx\paragraph\undefined\else
\let\oldparagraph\paragraph
\renewcommand{\paragraph}[1]{\oldparagraph{#1}\mbox{}}
\fi
\ifx\subparagraph\undefined\else
\let\oldsubparagraph\subparagraph
\renewcommand{\subparagraph}[1]{\oldsubparagraph{#1}\mbox{}}
\fi

\title{LightSAFT: Lightweight Latent Source Aware Frequency Transform
for Source Separation}

        \author[1]{Yeong-Seok Jeong\footnote{co-first author}}
          \author[1]{Jinsung Kim\footnote{co-first author}}
          \author[2]{Woosung Choi}
          \author[3]{Jaehwa Chung}
          \author[1]{Soonyoung Jung\footnote{corresponding author}}
    
      \affil[1]{Korea University}
      \affil[2]{Queen Mary University of London}
      \affil[3]{Korea National Open University}
  \date{\vspace{-7ex}}

\begin{document}
\maketitle

\marginpar{

  \begin{flushleft}
  \sffamily\small

    {\bfseries Arxiv DOI:} \href{https://doi.org/10.21105/joss.01667}{\color{linky}{10.21105/joss.01667}}
  
  \vspace{2mm}

  \par\noindent\hrulefill\par

  \vspace{2mm}

  \vspace{2mm}
  {\bfseries License}\\
  Authors of papers retain copyright and release the work under a Creative Commons Attribution 4.0 International License (\href{http://creativecommons.org/licenses/by/4.0/}{\color{linky}{CC BY 4.0}}).

  \vspace{4mm}
  {\bfseries In partnership with}\\
  \vspace{2mm}
  \includegraphics[width=4cm]{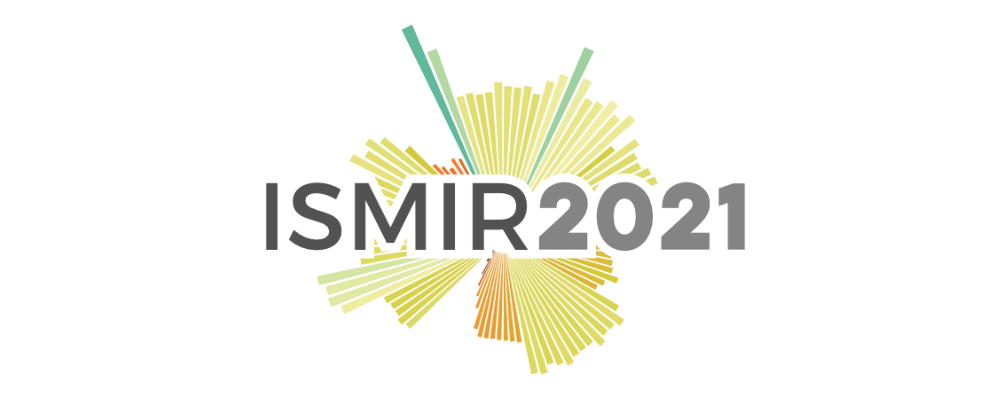}

  \end{flushleft}
}

\hypertarget{abstract}{%
\section{Abstract}\label{abstract}}

Conditioned source separations have attracted significant attention
because of their flexibility, applicability and extensionality. Their
performance was usually inferior to the existing approaches, such as the
single source separation model. However, a recently proposed method
called LaSAFT-Net has shown that conditioned models can show comparable
performance against existing single-source separation models. This paper
presents LightSAFT-Net, a lightweight version of LaSAFT-Net. As a
baseline, it provided a sufficient SDR performance for comparison during
the Music Demixing Challenge at ISMIR 2021. 

\hypertarget{introduction}{%
\section{Introduction}\label{introduction}}

Recently, many methods based on machine learning have been conducted for
music source separation. They can be distinguished depending upon
strategies for separation as follows.: single source separation (Choi et
al., 2019; Jansson et al., 2017; Takahashi \& Mitsufuji, 2017),
multi-head source separation (Défossez et al., 2019; Doire \& Okubadejo,
2019; Kadandale et al., 2020; Manilow, Seetharaman, et al., 2020; Sawata
et al., 2021), one-hot conditioned separation (Choi, Kim, Chung, et al.,
2021; Meseguer-Brocal \& Peeters, 2019; Samuel et al., 2020; Slizovskaia
et al., 2021), Query-by-Example (QBE) separation (Lee et al., 2019; Lin
et al., 2021), recursive separation (Takahashi et al., 2019; Wichern et
al., 2019), embedding space separation Luo et al. (2017) and
hierarchical separation (Manilow, Wichern, et al., 2020). The
conditioned source separation method separates target source from a
given mixture depending on its conditioning input, which can be various
forms (e.g., one-hot conditions, QBE, etc.). Although it has to separate
with more complex mechanisms than the multi-head separation method, it
is still promising because of its flexibility, applicability and
extensionality. For example, it's easy to extend another domain such as
text-conditioned audio manipulation (Choi, Kim, Mart{\'\i}nez Ram{\'\i}rez, et al., 2021).
In this paper, we focus on the one-hot conditioned source separation
method since it has extensionality.

LaSAFT-Net (Choi, Kim, Chung, et al., 2021) is one of the representative
conditioned source separation models. It applies the Latent Source aware
Frequency Transformation (LaSAFT) blocks. A LaSAFT block aims to capture
the latent source's frequency patterns depending upon a given symbol
that specifies which target source we want to separate. Assuming that
each latent source contains independent information depending on its
viewpoint, it employs the attention mechanism to model the relevance
between latent sources and the target symbol. With this approach,
LaSAFT-Net showed comparable performance to single-source separation
models on the MusDB18 (Rafii et al., 2017) benchmark even if conditioned
models are usually inferior to single-source separation models.

However, it consumes numerous parameters to analyze latent sources. It
prevents the model's applicability to a real-world environment, where
resources are restricted, such as Music Demix Challenge (MDX) (Mitsufuji
et al., 2021) at ISMIR 2021. To this end, we explore a lightweight
version of LaSAFT-Net to make it affordable in a real-world environment
by decreasing the number of parameters and simplifying the structure.

In this paper, we explore the method for the light version of
LaSAFT-Net. The existing LaSAFT-Net uses numerous parameters to separate
each latent source. We focus on reducing parameters by introducing
shared components in latent source analysis. 
Through this method, the LightSAFT-Net is affordable in a restricted
environment and shows sufficient performance in MDX challenge.

\hypertarget{related-work}{%
\section{Related work}\label{related-work}}

\hypertarget{latent-source}{%
\subsection{Latent Source}\label{latent-source}}

Conditioned music source separation models, such as LaSAFT-Net (Choi,
Kim, Chung, et al., 2021) and Meta-TasNet (Samuel et al., 2020), aim to
separate symbolic-labeled target instruments sets defined by the
training dataset. However, some musical sub-components can be grouped as
a single instrument even if they have different sonic characteristics in
such dataset. For example, bass-drum and Hi-hat are included in `drums
set' though they have different timbre in the MUSDB18 dataset (Rafii et
al., 2017). Arbitrary instruments set, which do not share the common
acoustic features, might make a model hinder separating it from a
mixture. It is even more challenging for conditioned source separation
models, which modulate internal features according to a given
conditioning symbol.

Some existing studies adopted latent source analysis (Choi, Kim, Chung,
et al., 2021; Choi, Kim, Mart{\'\i}nez Ram{\'\i}rez, et al., 2021; Wisdom et al., 2020) to
relax the problems caused by symbolically labeled instruments' sets.
They assume that a model can learn a latent source that can deal with a
more detailed aspect of sound from a given training dataset. For
example, a LaSAFT block aims to analyze frequency patterns of latent
sources. It aggregates the analyzed patterns by computing an attention
matrix between latent sources and a given input conditioning symbol
(e.g., one-hot encoding vector that specifies which source a user wants
to separate). In this paper, we adopt the latent source analysis method,
proposed in (Choi, Kim, Chung, et al., 2021), to analyze sub-components
of grouped instrument's set.

\hypertarget{light-weight-latent-source-attentive-frequency-transformation-network}{%
\section{Light-weight latent Source Attentive Frequency Transformation
Network}\label{light-weight-latent-source-attentive-frequency-transformation-network}}

\hypertarget{latent-source-attentive-separation-mechanism}{%
\subsection{Latent Source Attentive Separation
mechanism}\label{latent-source-attentive-separation-mechanism}}

We present how the original LaSAFT block processes an input feature as
follows. A LaSAFT block analyzes frequency patterns of latent sources
using Time Distributed Fully connected layer (TDF) blocks and aggregates the analyzed patterns by computing an
attention matrix between latent sources and a given input conditioning
symbol. An attention score measures how important the relationship
between latent sources and the target is for separation. To compute an
attention matrix, it first generates a query
vector(\(Q \in \mathbb{R}^{1 \times d_{k}}\)), which has \(d_{k}\)
dimensions, using an embedding layer. Key
vectors(\(K \in \mathbb{R}^{|S_L| \times d_{k}}\)) are representative
vectors of latent sources, which are learnable parameters. To compute
the attention scores, we apply a dot product of the query and key and
take softmax after scaled by \(\sqrt{d_k}\).

The multiple TDF blocks generate latent source features
\(V \in \mathbb{R}^{T \times d_{k} \times |S_L|}\), where the \(T\) is
the number of frames. The attention mechanism which mixes the latent
sources' features is as follows:
\[Attention(Q,K,V)=softmax(QK^{T}/\sqrt{d_{k}})V\]

\hypertarget{lightweight-latent-source-attentive-frequency-transformation-block}{%
\subsection{Lightweight latent Source Attentive Frequency Transformation
Block}\label{lightweight-latent-source-attentive-frequency-transformation-block}}

The TDF of the original LaSAFT block contains two phases (Figure 1.(a)).
In the first phase, fully connected layers extract latent sources'
features from mixture sources. In the second phase, other linear layers
are applied to generate the final output. Even if each fully connected
layers downsample the internal space, this architecture still consumes
many parameters. It seems that it contains too many connections between
two phases because a connection exists for each pair of different latent
sources. Although such connections might be helpful for separation
enabling active communication between sub-modules, it may be
computationally difficult to use this model in real-time environment.
Therefore, we focus on this point and explore the methods for lightening
the model's parameters and maintaining the performance.

\begin{figure}
\centering
\includegraphics{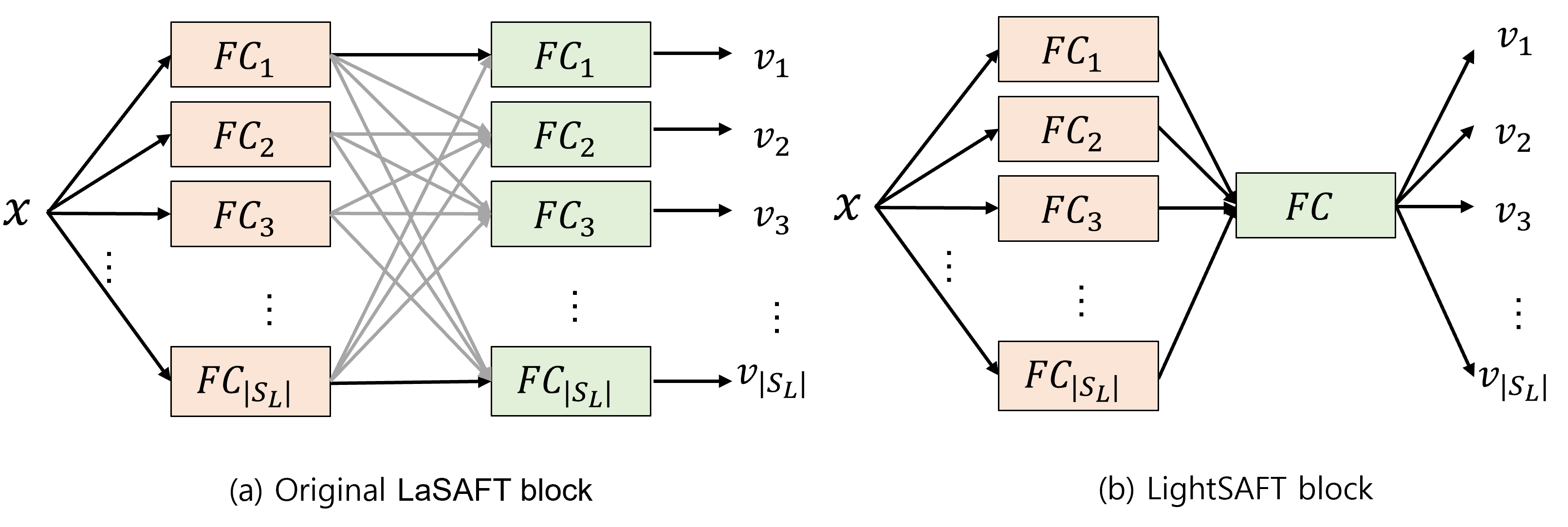}
\caption{Comparison the original LaSAFT block and proposed
LightSAFT block. The $x$ is the input intermediate feature
and the $v_i$ is generated latent source.}
\end{figure}

Figure 1 shows the difference between the original LaSAFT block and the
proposed LightSAFT block in the latent source separating process. The
blocks receive the intermediate feature x and generate the latent source
V. Each FC block comprises a fully connected layer (FC), Batch Norm, and
activation function in each block. We eliminate inter-source connections
between two phases and use a shared linear layer in the in phase 2. The
LightSAFT block has only one separating process, assuming that the
redundant separating process in the original block is unnecessary. From
this approach, we improve the network's applicability and efficiency. We
evaluated the proposed models' performance on the MDX benchmark system
(Mitsufuji et al., 2021), which did not allow GPU computing for
acceleration. While the original LaSAFT-Net could not be evaluated
within the time limit on the system, the proposed models do not exceed
the time limit. We summarize the evaluation results in Section
``Experiments.''

\hypertarget{experiments}{%
\section{Experiments}\label{experiments}}

\hypertarget{model-configuration}{%
\subsection{Model Configuration}\label{model-configuration}}

To fairly compare the number of parameters and performance, we follow
the configurations of Choi, Kim, Chung, et al. (2021). The \(|S_L|\) is
16 and \(d_k\) is 24. Those are the same as the original LaSAFT-Net's
configuration (\$ Table 1). It is less than half of the number of the
LaSAFT-Net's parameters. We trained our proposed model by minimizing the
mean squared error between the target's short-Time Fourier
Transformation (STFT) and the model's output.\\
For data augmentation, we generated the mixtures by mixing the different
track's sources.

\hypertarget{results}{%
\subsection{Results}\label{results}}

\begin{longtable}[]{@{}llllllll@{}}
\toprule
model & Runnable? & \# of parameters & vocals & drums & bass & other &
Avg\tabularnewline
\midrule
\endhead
LaSAFT-Net & no & 4.5M & - & - & - & - & -\tabularnewline
LightSAFT-Net & yes & 3.8M & 6.685 & 5.272 & 5.498 & 4.121 &
5.394\tabularnewline
\bottomrule
\end{longtable}

\textbf{Table 1}. A comparison with LaSAFT-Net. The \emph{Runnable?}
means that the model is runnable in MDX environment.

\begin{longtable}[]{@{}lllllll@{}}
\toprule
model & type & vocals & drums & bass & other & Avg\tabularnewline
\midrule
\endhead
Demucs48-HQ (Défossez et al., 2019) & Single & 6.496 & 6.509 & 6.470 &
4.018 & 5.873\tabularnewline
XUMX (Sawata et al., 2021) & multi-head & 6.341 & 5.807 & 5.615 & 3.722
& 5.372\tabularnewline
UMX (Stöter et al., 2019) & Single & 5.999 & 5.504 & 5.357 & 3.309 &
5.042\tabularnewline
LightSAFT-Net & conditioned & 6.685 & 5.272 & 5.498 & 4.121 &
5.394\tabularnewline
\bottomrule
\end{longtable}

\textbf{Table 2}. A comparison with other source separation models

We compare the performance between the original model and the proposed
models in the same condition.\\
Table 1 shows the results of the model's SDR (Vincent et al., 2006)
score in the MDX challenge and the number of each model's parameters. In
same condition, the original LaSAFT-Net's parameters are 4.5M, while the
LightSAFT-Net has 3.8M parameters sufficient compression for the MDX
challenge.
The LaSAFT-Net, which cannot separate the songs
in a limited time, was not reasonable for this challenge.
On the other hand, the LightSAFT-Net, which is posted as a comparison of
the MDX challenge, can separate the music source in a limited time and
achieve comparable performance.

Usually, the conditioned source separation models, which can separate
all kinds of sources, show inferior performance than the single source
separation model; the conditioned model can not focus on the specific
task since it have to learn generalized weights for conducting all tasks
in limited model complexity. Despite performance degradation, the
conditioned source separation model is more attractive because of its
applicability and efficiency. Table 2 shows whether the model is
conditioned or not and its performance. We take other models'
performance at the MDX Leaderboard A, which do not allow additional
datasets, for precise comparison.\\
The LightSAFT-Net shows competitive performance despite the conditioned
source separation model. Even the model shows a better performance than
Demucs-HQ, another baseline of the MDX challenge.

\hypertarget{conclusion}{%
\section{Conclusion}\label{conclusion}}

We explore the method to reduce the number of the model's parameters and
maintain the performance. We show that our method is reasonable for
source separation tasks in the performance and the applicability, which
can be used in a restricted environment like the MDX challenge.\\
The LightSAFT-Net shows the competitive performance even though it is a
conditioned source separation model. For future work, we will study the
method to extend it to other tasks.

\hypertarget{acknowledgment}{%
\section{Acknowledgment}\label{acknowledgment}}

This research was supported by Basic Science Research Program through
the National Research Foundation of Korea(NRF) funded by the Ministry of
Education(NRF-2021R1A6A3A03046770). This work was also supported by the
National Research Foundation of Korea(NRF) grant funded by the Korea
government(MSIT)(No.~NRF-2020R1A2C1012624, NRF-2021R1A2C2011452).

\hypertarget{reference}{%
\section*{Reference}\label{reference}}
\addcontentsline{toc}{section}{Reference}

\hypertarget{refs}{}
\begin{CSLReferences}{1}{0}
\leavevmode\hypertarget{ref-choi:2020}{}%
Choi, W., Kim, M., Chung, J., \& Jung, S. (2021). LaSAFT: Latent source
attentive frequency transformation for conditioned source separation.
\emph{ICASSP 2021-2021 IEEE International Conference on Acoustics,
Speech and Signal Processing (ICASSP)}, 171--175.

\leavevmode\hypertarget{ref-choi:2019}{}%
Choi, W., Kim, M., Chung, J., Lee, D., \& Jung, S. (2019). Investigating
deep neural transformations for spectrogram-based musical source
separation. \emph{arXiv Preprint arXiv:1912.02591}.

\leavevmode\hypertarget{ref-choi:2021}{}%
Choi, W., Kim, M., Ram{\'\i}rez, M. A. M., Chung, J., \& Jung, S. (2021).
AMSS-net: Audio manipulation on user-specified sources with textual
queries. \emph{arXiv Preprint arXiv:2104.13553}.

\leavevmode\hypertarget{ref-defossez:2019}{}%
Défossez, A., Usunier, N., Bottou, L., \& Bach, F. (2019). Music source
separation in the waveform domain. \emph{arXiv Preprint
arXiv:1911.13254}.

\leavevmode\hypertarget{ref-doire:2019}{}%
Doire, C. S., \& Okubadejo, O. (2019). Interleaved multitask learning
for audio source separation with independent databases. \emph{arXiv
Preprint arXiv:1908.05182}.

\leavevmode\hypertarget{ref-jansson:2017}{}%
Jansson, A., Humphrey, E., Montecchio, N., Bittner, R., Kumar, A., \&
Weyde, T. (2017). \emph{Singing voice separation with deep u-net
convolutional networks}.

\leavevmode\hypertarget{ref-kadandale:2020}{}%
Kadandale, V. S., Montesinos, J. F., Haro, G., \& Gómez, E. (2020).
\emph{Multi-task u-net for music source separation}.

\leavevmode\hypertarget{ref-lee:2019}{}%
Lee, J. H., Choi, H.-S., \& Lee, K. (2019). Audio query-based music
source separation. In A. Flexer, G. Peeters, J. Urbano, \& A. Volk
(Eds.), \emph{Proceedings of the 20th international society for music
information retrieval conference, {ISMIR} 2019, delft, the netherlands,
november 4-8, 2019} (pp. 878--885).
\url{http://archives.ismir.net/ismir2019/paper/000108.pdf}

\leavevmode\hypertarget{ref-lee:2019}{}%
Lee, J. H., Choi, H.-S., \& Lee, K. (2019). Audio query-based music
source separation. In A. Flexer, G. Peeters, J. Urbano, \& A. Volk
(Eds.), \emph{Proceedings of the 20th international society for music
information retrieval conference, {ISMIR} 2019, delft, the netherlands,
november 4-8, 2019} (pp. 878--885).
\url{http://archives.ismir.net/ismir2019/paper/000108.pdf}

\leavevmode\hypertarget{ref-lin:2021}{}%
Lin, L., Kong, Q., Jiang, J., \& Xia, G. (2021). A unified model for
zero-shot music source separation, transcription and synthesis.
\emph{arXiv Preprint arXiv:2108.03456}.

\leavevmode\hypertarget{ref-luo:2017}{}%
Luo, Y., Chen, Z., Hershey, J. R., Le Roux, J., \& Mesgarani, N. (2017).
Deep clustering and conventional networks for music separation: Stronger
together. \emph{2017 IEEE International Conference on Acoustics, Speech
and Signal Processing (ICASSP)}, 61--65.

\leavevmode\hypertarget{ref-manilow:2020}{}%
Manilow, E., Seetharaman, P., \& Pardo, B. (2020). Simultaneous
separation and transcription of mixtures with multiple polyphonic and
percussive instruments. \emph{ICASSP 2020-2020 IEEE International
Conference on Acoustics, Speech and Signal Processing (ICASSP)},
771--775.

\leavevmode\hypertarget{ref-ethan:2020}{}%
Manilow, E., Wichern, G., \& Le Roux, J. (2020). Hierarchical musical
instrument separation. \emph{International Society for Music Information
Retrieval (ISMIR) Conference}, 376--383.
ISBN:~\href{https://worldcat.org/isbn/978-0-9813537-0-8}{978-0-9813537-0-8}

\leavevmode\hypertarget{ref-cunet:2019}{}%
Meseguer-Brocal, G., \& Peeters, G. (2019). CONDITIONED-u-NET:
INTRODUCING a CONTROL MECHANISM IN THE u-NET FOR MULTIPLE SOURCE
SEPARATIONS. \emph{Proceedings of the 20th International Society for
Music Information Retrieval Conference}.

\leavevmode\hypertarget{ref-cunet:2019}{}%
Meseguer-Brocal, G., \& Peeters, G. (2019). CONDITIONED-u-NET:
INTRODUCING a CONTROL MECHANISM IN THE u-NET FOR MULTIPLE SOURCE
SEPARATIONS. \emph{Proceedings of the 20th International Society for
Music Information Retrieval Conference}.

\leavevmode\hypertarget{ref-mitsufuji:2021}{}%
Mitsufuji, Y., Fabbro, G., Uhlich, S., \& Stöter, F.-R. (2021). Music
demixing challenge at ISMIR 2021. \emph{arXiv Preprint
arXiv:2108.13559}.

\leavevmode\hypertarget{ref-rafii:2017}{}%
Rafii, Z., Liutkus, A., Stöter, F.-R., Mimilakis, S. I., \& Bittner, R.
(2017). \emph{MUSDB18-a corpus for music separation}.

\leavevmode\hypertarget{ref-samuel:2020}{}%
Samuel, D., Ganeshan, A., \& Naradowsky, J. (2020). Meta-learning
extractors for music source separation. \emph{ICASSP 2020-2020 IEEE
International Conference on Acoustics, Speech and Signal Processing
(ICASSP)}, 816--820.

\leavevmode\hypertarget{ref-samuel:2020}{}%
Samuel, D., Ganeshan, A., \& Naradowsky, J. (2020). Meta-learning
extractors for music source separation. \emph{ICASSP 2020-2020 IEEE
International Conference on Acoustics, Speech and Signal Processing
(ICASSP)}, 816--820.

\leavevmode\hypertarget{ref-sawata:2021}{}%
Sawata, R., Uhlich, S., Takahashi, S., \& Mitsufuji, Y. (2021). All for
one and one for all: Improving music separation by bridging networks.
\emph{ICASSP 2021-2021 IEEE International Conference on Acoustics,
Speech and Signal Processing (ICASSP)}, 51--55.

\leavevmode\hypertarget{ref-seetharaman:2019}{}%
Seetharaman, P., Wichern, G., Venkataramani, S., \& Le Roux, J. (2019).
Class-conditional embeddings for music source separation. \emph{ICASSP
2019-2019 IEEE International Conference on Acoustics, Speech and Signal
Processing (ICASSP)}, 301--305.

\leavevmode\hypertarget{ref-olga:2021}{}%
Slizovskaia, O., Haro, G., \& Gómez, E. (2021). Conditioned source
separation for musical instrument performances. \emph{IEEE/ACM
Transactions on Audio, Speech, and Language Processing}, \emph{29},
2083--2095.

\leavevmode\hypertarget{ref-stoter:2019}{}%
Stöter, F.-R., Uhlich, S., Liutkus, A., \& Mitsufuji, Y. (2019).
Open-unmix-a reference implementation for music source separation.
\emph{Journal of Open Source Software}, \emph{4}(41), 1667.

\leavevmode\hypertarget{ref-takahashi:2017}{}%
Takahashi, N., \& Mitsufuji, Y. (2017). \emph{Multi-scale multi-band
DenseNets for audio source separation}.
\url{http://arxiv.org/abs/1706.09588}

\leavevmode\hypertarget{ref-recursive:2019}{}%
Takahashi, N., Parthasaarathy, S., Goswami, N., \& Mitsufuji, Y. (2019).
Recursive speech separation for unknown number of speakers. In G. Kubin
\& Z. Kacic (Eds.), \emph{Interspeech 2019, 20th annual conference of
the international speech communication association, graz, austria, 15-19
september 2019} (pp. 1348--1352). {ISCA}.
\url{https://doi.org/10.21437/Interspeech.2019-1550}

\leavevmode\hypertarget{ref-vincent:2006}{}%
Vincent, E., Gribonval, R., \& Févotte, C. (2006). Performance
measurement in blind audio source separation. \emph{IEEE Transactions on
Audio, Speech, and Language Processing}, \emph{14}(4), 1462--1469.

\leavevmode\hypertarget{ref-wichern:2019}{}%
Wichern, G., Antognini, J., Flynn, M., Zhu, L. R., McQuinn, E., Crow,
D., Manilow, E., \& Roux, J. L. (2019). WHAM!: Extending speech
separation to noisy environments. \emph{arXiv Preprint
arXiv:1907.01160}.

\leavevmode\hypertarget{ref-wisdom:2020}{}%
Wisdom, S., Tzinis, E., Erdogan, H., Weiss, R. J., Wilson, K., \&
Hershey, J. R. (2020). Unsupervised sound separation using mixture
invariant training. \emph{NeurIPS}.
\url{https://arxiv.org/pdf/2006.12701.pdf}

\end{CSLReferences}

\end{document}